\documentclass[a4paper]{jacow}

\usepackage{pdfpages,multirow,ragged2e}

\makeatletter%
	\ifboolexpr{bool{xetex}}
	 {\renewcommand{\Gin@extensions}{.pdf,%
	                    .png,.jpg,.bmp,.pict,.tif,.psd,.mac,.sga,.tga,.gif,%
	                    .eps,.ps,%
	                    }}{}
\makeatother

\ifboolexpr{bool{xetex} or bool{luatex}}
 {}
 {\usepackage[utf8]{inputenc}}

\usepackage[USenglish]{babel}

\begin{document}

\title{Updates on Impedance Studies for the FCC-ee High Energy Booster}

\author{
K. H. Kim\thanks{keon-hee.kim@ganil.fr}, A. Ghribi\thanks{adnan.ghribi@cern.ch}\textsuperscript{1}, S. Martinez\textsuperscript{2},
GANIL, Caen, France \\
Q. Bruant, A. Chance, B. Dalena, CEA, Gif-sur-Yvette, France \\
C. Antuono, D. Gibellieri\textsuperscript{1}, C. Zannini, F. Zimmermann, CERN, Geneva, Switzerland \\
A. Mashal, IPM, Tehran, Iran \\
M. Migliorati, University of Rome La Sapienza and INFN - Roma1, Rome, Italy \\
M. Zobov, INFN - LNF, Frascati, Rome, Italy \\
\textsuperscript{1}also at University of Caen Normandy, Caen, France \\
\textsuperscript{2}also at Paris-Saclay University, Orsay, France \\
}

\maketitle

\begin{abstract}
Following the Future Circular Collider (FCC) Feasibility Study completion, the impedance model for the FCC-ee High-Energy Booster (HEB) has been significantly expanded beyond the initial copper vacuum pipe resistive wall analysis. This paper presents a comprehensive impedance and wake budget incorporating RF cavities, bellows, and beam position monitors, evaluated through 3D electromagnetic simulations and analytical methods.

The updated model provides the basis for future beam dynamics studies, including transverse coupled bunch instability analyses and single bunch tracking simulations. The present work focuses on the construction and comparison of the main impedance and wake contributions, identifying the dominant sources and the components requiring further investigation. These results will be used to refine the HEB collective effects studies and to support future assessments of instability margins and mitigation requirements.
\end{abstract}

\section{INTRODUCTION}

The FCC-ee HEB is an essential part of the FCC-ee injector complex. It accelerates the beams from the injection energy to the collider injection energy. Therefore, its collective effects must be carefully assessed over the relevant operating range. In particular, the impedance and wake model of the HEB provides the input for single bunch tracking studies, transverse mode coupling instability scans, and coupled bunch instability analyses.

Previous studies of the HEB mainly focused on the resistive wall contribution of the vacuum chamber~\cite{gibellieri-ipac2025}. In the present work, the impedance model is extended to include additional geometric impedance sources: bellows, beam position monitors (BPMs), and RF cavities. The resistive wall contribution is evaluated analytically using ImpedanceWake2D (IW2D)~\cite{iw2d}, while the geometric contributions are obtained from 3D electromagnetic simulations performed with CST Studio Suite~\cite{cst-studio-suite}.

The present study focuses on the Z operation mode, with an injection energy of \SI{20}{GeV} and an extraction energy of \SI{45.6}{GeV}. This configuration is particularly relevant for impedance studies because of its small emittance and sensitivity to collective effects. The goal of this paper is to present the updated longitudinal and transverse impedance budgets and the corresponding wake potentials for a short bunch.

\section{HEB MACHINE PARAMETERS}

The main machine parameters used in this study are summarized in Table~\ref{tab:heb_machine_parameters}. The parameters correspond to the Z operation mode.

\begin{table}[!htb]
   \centering
   \caption{Machine Parameters of the HEB for Z Operation}
   \label{tab:heb_machine_parameters}
   \footnotesize
   \begin{tabular}{lc}
       \toprule
       \textbf{Parameter} & \textbf{Value} \\
       \midrule
       Circumference (km) & 90.644836 \\
       Injection beam energy (GeV) & 20 \\
       Extraction beam energy (GeV) & 45.6 \\
       Bunch population, filling ($10^{11}$) & 0.25 \\
       Bunch population, top-up ($10^{11}$) & 0.216 \\
       RF frequency (MHz) & 800 \\
       Injection RF voltage (MV) & 64.25 \\
       Extraction RF voltage (MV) & 60.98 \\
       Injection energy loss per turn (MeV/turn) & 1.31 \\
       Extraction energy loss per turn (MeV/turn) & 35.45 \\
       Longitudinal damping time at injection (s) & 4.61 \\
       Longitudinal damping time at extraction (s) & 0.389 \\
       Momentum compaction factor ($10^{-6}$) & 9.24 \\
       Horizontal tune & 368.225 \\
       Vertical tune & 283.29 \\
       Injection normalized emittance, Hor/Vert ($\mu$m) & 20/2 \\
       Injection RMS bunch length (mm) & 4 \\
       Injection RMS energy spread (\%) & 0.1 \\
       Extraction RMS bunch length (mm) & 2.62 \\
       Extraction RMS energy spread (\%) & 0.0386 \\
       \bottomrule
   \end{tabular}
\end{table}

\section{IMPEDANCE AND WAKE MODEL}

The impedance and wake model considered in this work is summarized in Table~\ref{tab:heb_impedance_model}. The model includes the resistive wall contribution of the beam pipe and the geometric contributions from bellows, BPMs, and RF cavities.

\begin{table}[!htb]
   \centering
   \caption{Evaluation Methods Used for the HEB Impedance and Wake Model}
   \label{tab:heb_impedance_model}
   \footnotesize
   \begin{tabular}{lcc}
       \toprule
       \textbf{Element} & \textbf{Number} & \textbf{Evaluation method} \\
       \midrule
       Beam pipe (RW) & 1 ring & Analytical (IW2D) \\
       Bellows & 10000 & 3D simulations (CST) \\
       BPMs & 2720 & 3D simulations (CST) \\
       RF cavities & 112 & 3D simulations (CST) \\
       \bottomrule
   \end{tabular}
\end{table}

The HEB beam pipe is modeled as a circular multilayer chamber with a radius of \SI{30}{mm}, including a \SI{150}{nm} NEG coating, a \SI{2}{mm} copper layer, a high-resistivity dielectric layer, and an outer stainless steel layer. In contrast to the main ring chamber, no winglets are included in the HEB baseline beam pipe.

The bellows impedance is adopted from the main ring design, assuming the same bellows geometry. The BPMs are based on the button type design used for the main ring, but with the HEB circular beam pipe as the baseline chamber. The RF contribution corresponds to the 800 MHz RF cryomodules, with 28 cryomodules and four cavities per cryomodule, giving 112 cavities in total. In the present booster design, no RF tapers are included.

For the transverse impedance budget, each localized contribution is weighted by the beta function at its location:
\begin{equation}
    Z_{\perp}(f)
    =
    \sum_{i=1}^{N}
    Z_{\perp,i}(f)
    \frac{\beta_{\perp}^{i}}{\langle \beta_{\perp} \rangle}.
    \label{eq:beta_weighting}
\end{equation}
Here, $Z_{\perp,i}(f)$ is the transverse impedance of the $i$-th element, $\beta_{\perp}^{i}$ is the beta function at the location of this element, and $\langle \beta_{\perp} \rangle$ is the average beta function in the corresponding transverse plane. This weighting is applied separately in the horizontal and vertical planes.

\section{TOTAL IMPEDANCE BUDGET}

\begin{figure}[!htb]
    \centering
    \includegraphics[width=1.0\columnwidth]{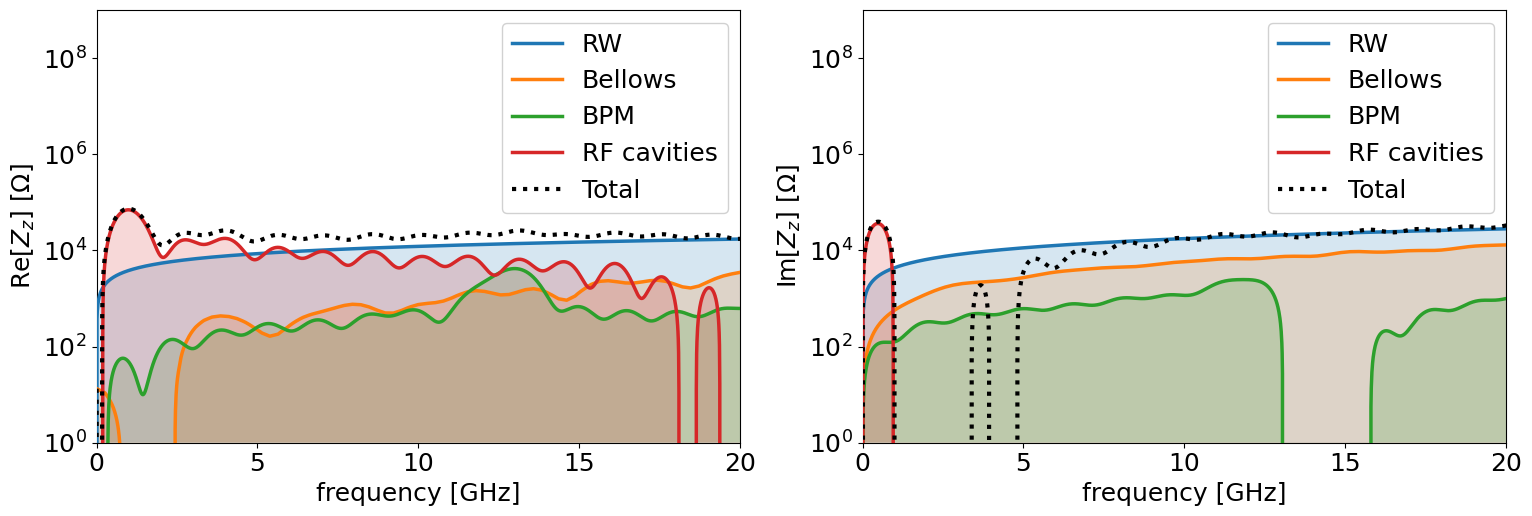}
    \caption{Longitudinal impedance budget of the HEB.}
    \label{fig:long_impedance}
\end{figure}

The resulting longitudinal impedance budget is shown in Fig.~\ref{fig:long_impedance}. The total longitudinal impedance is obtained by combining the resistive wall contribution with the geometric contributions from the bellows, BPMs, and RF cavities.

\begin{figure}[!htb]
    \centering
    \includegraphics[width=1.0\columnwidth]{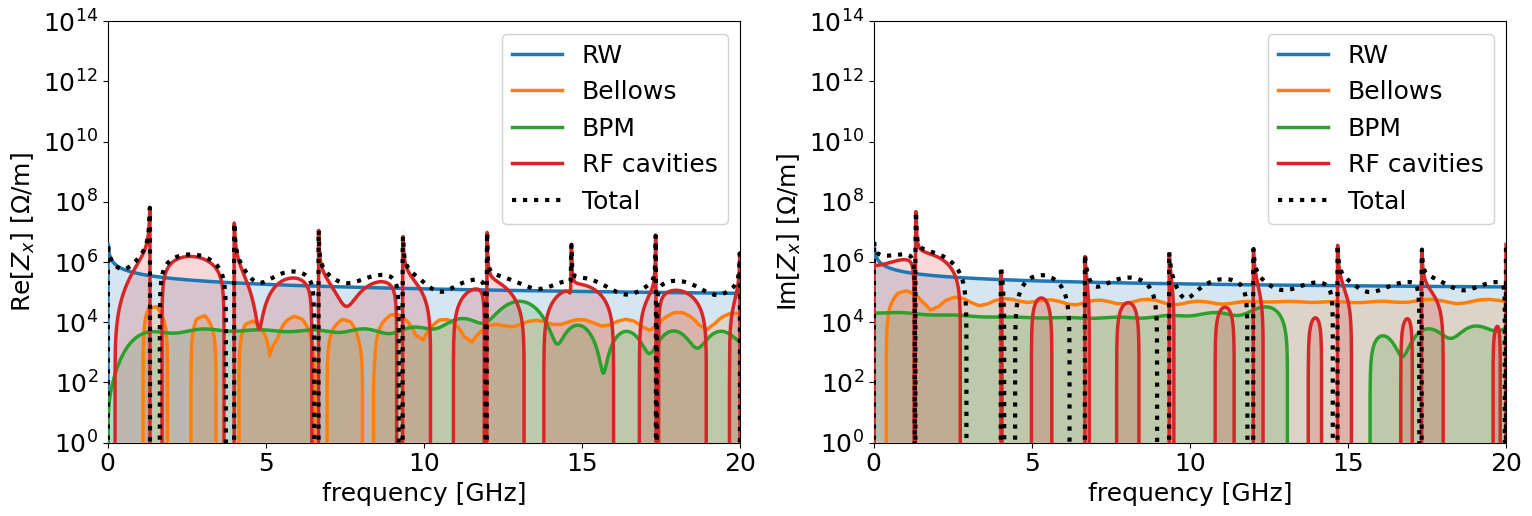}
    \caption{Horizontal dipolar impedance budget of the HEB.}
    \label{fig:dipx_impedance}
\end{figure}

\begin{figure}[!htb]
    \centering
    \includegraphics[width=1.0\columnwidth]{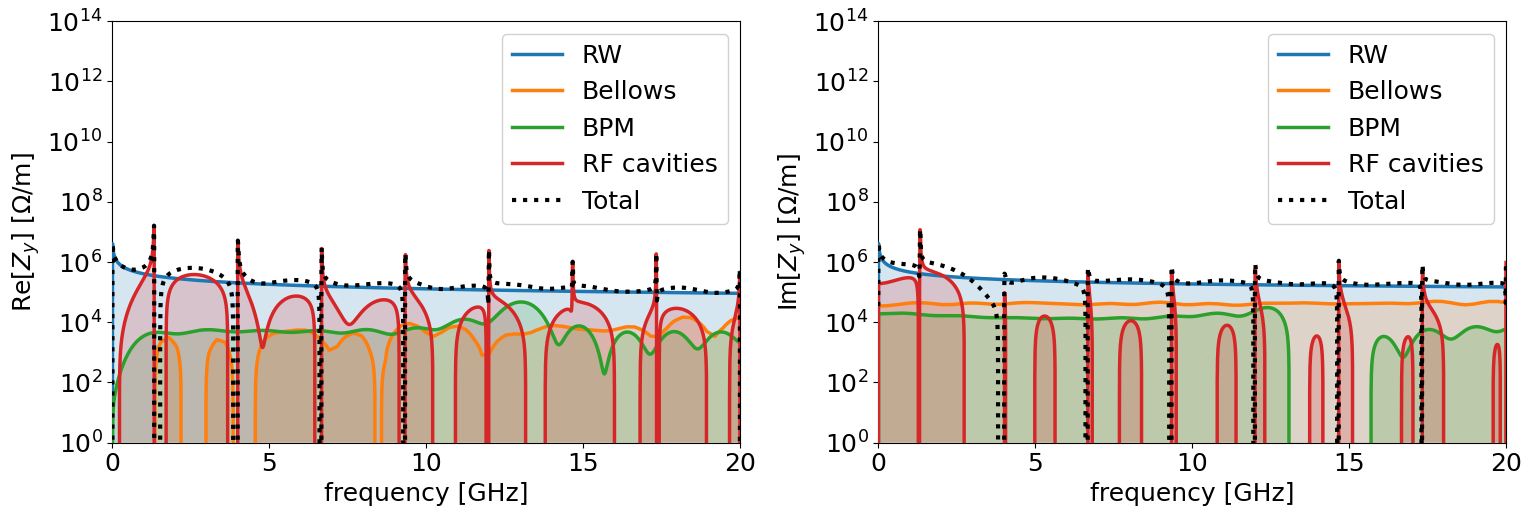}
    \caption{Vertical dipolar impedance budget of the HEB.}
    \label{fig:dipy_impedance}
\end{figure}

The corresponding horizontal and vertical dipolar impedance budgets are shown in Figs.~\ref{fig:dipx_impedance} and~\ref{fig:dipy_impedance}. The beta weighting procedure described in Eq.~\eqref{eq:beta_weighting} is applied to the localized transverse impedance contributions.

The present HEB impedance budget is largely driven by the resistive wall contribution of the circular beam pipe, while the RF cavity contribution is also non-negligible. This is still a favorable outcome, since the resistive wall impedance is essentially unavoidable and the bellows and BPM contributions remain moderate. The relative importance of the RF contribution motivates a more detailed assessment in subsequent beam dynamics studies.

\section{WAKE POTENTIALS}

Wake potentials were evaluated for an RMS bunch length of \SI{0.4}{mm}. This bunch length is sufficiently short to be used in the pseudo-Green function approach, as demonstrated in Ref.~\cite{migliorati-epjp2021}, and provides suitable wake inputs for future beam dynamics simulations. It also allows the relative importance of the different impedance sources to be compared in the short-range regime.

\begin{figure}[!htb]
    \centering
    \includegraphics[width=0.80\columnwidth]{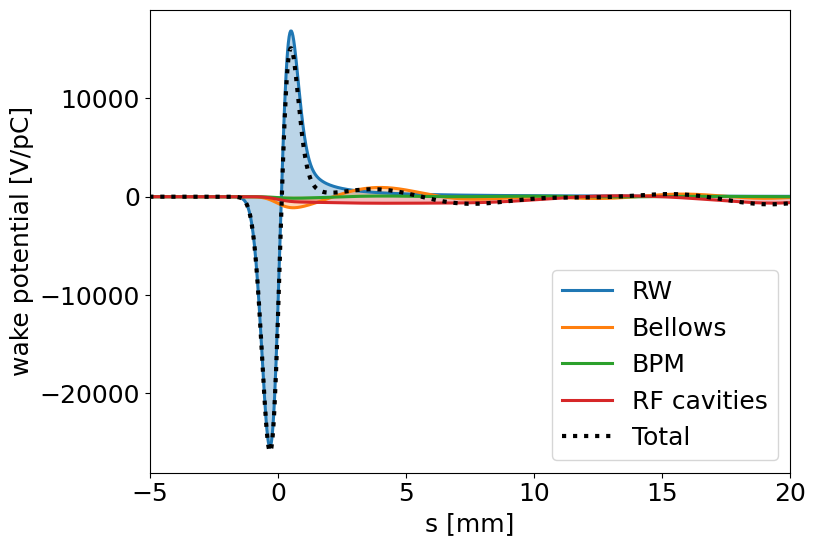}
    \caption{Longitudinal wake potential for an RMS bunch length of \SI{0.4}{mm}.}
    \label{fig:long_wake}
\end{figure}

\begin{figure}[!htb]
    \centering
    \includegraphics[width=1.0\columnwidth]{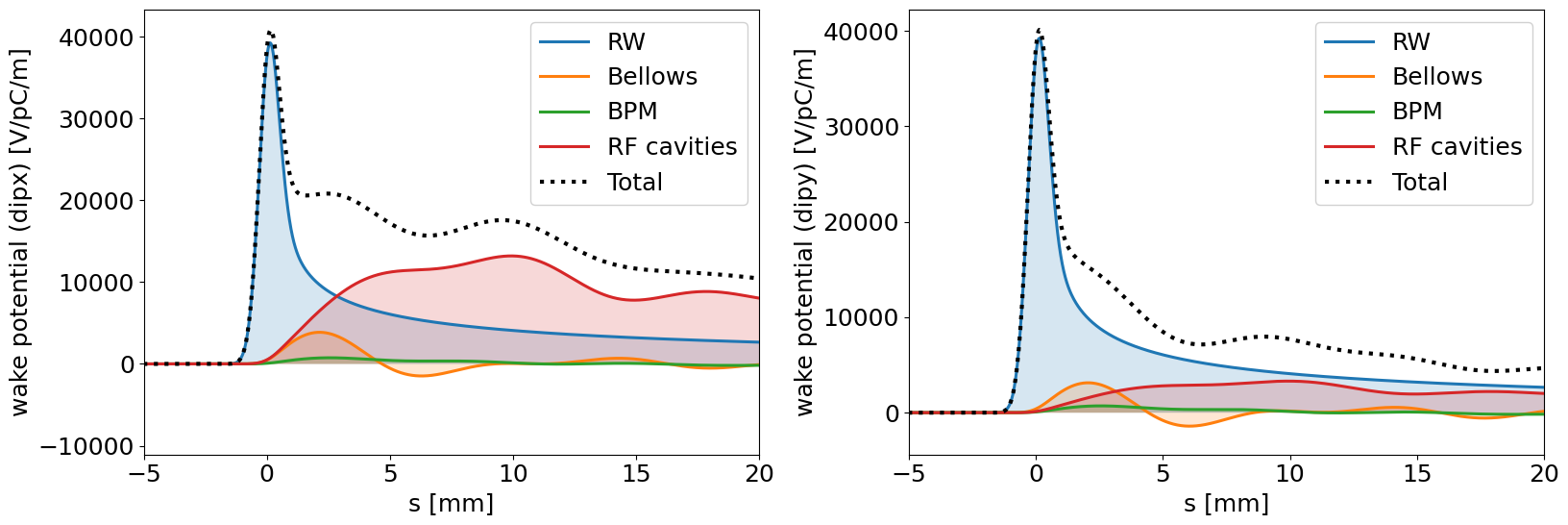}
    \caption{Horizontal and vertical dipolar wake potentials for an RMS bunch length of \SI{0.4}{mm}.}
    \label{fig:trans_wake}
\end{figure}

The longitudinal wake potential is shown in Fig.~\ref{fig:long_wake}. The transverse dipolar wake potentials are shown in Fig.~\ref{fig:trans_wake}.

The geometric contributions from the bellows and BPMs remain moderate in the present budget. In particular, the BPM contribution benefits from the smaller number of BPMs in the HEB compared with the corresponding main ring model~\cite{gibellieri-ipac2025}.

The RF cavity contribution nevertheless requires particular attention. Although the HEB RF system does not include RF tapers in the present design, the RF cavity contribution is found to be significantly larger than in the corresponding main ring case~\cite{gibellieri-ipac2025}. Its impact on the longitudinal and transverse beam dynamics is therefore under detailed investigation.

\section{FUTURE DIRECTIONS}

\subsection{Beam Dynamics Studies}

The present work focuses on the construction of the impedance and wake budget for the FCC-ee HEB. The next step is to use this updated model as an input for beam dynamics simulations, in order to quantify its impact on both longitudinal and transverse collective effects. Single bunch tracking studies, transverse mode coupling instability scans, and coupled bunch instability analyses will be performed for the relevant HEB operating conditions, including injection and extraction energies. These studies will allow us to identify the most critical impedance sources and to assess whether additional design optimization or mitigation measures are required. Special attention will be given to the RF cavity contribution, whose relative impact appears larger than expected from the corresponding main ring studies.

\subsection{Neural Operators}

As a longer-term direction, neural operators~\cite{kovachki2023neural} will be investigated as surrogate models for longitudinal collective effects studies. In particular, the goal is to approximate the collective beam evolution induced by the updated wake model, enabling fast scans over bunch intensity, RF parameters, impedance scaling, and machine configurations. Such an approach could complement conventional tracking simulations by providing a fast reduced model once trained on high fidelity beam dynamics data.

\begin{figure}[!htb]
\centering
\includegraphics*[width=1.0\columnwidth]{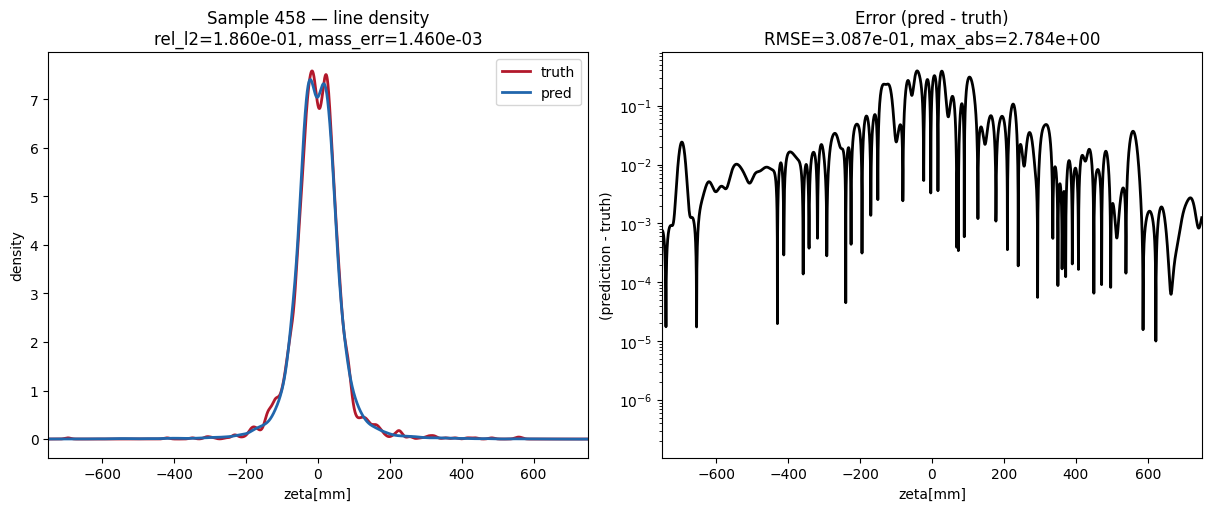}
\caption{Representative evolution of the longitudinal beam distribution through a section of the HEB lattice, illustrating the non-local collective transport dynamics learned by the Fourier neural operator.}
\label{fig:fno1d}
\end{figure}

The beam dynamics are governed by the Vlasov--Fokker--Planck equation
\begin{equation}
    \frac{\partial \rho}{\partial s} + \{\rho,H\} = \mathcal{C}[\rho],
\end{equation}
whose formal solution may be written as 
\begin{equation}
    \rho_{s+\Delta s}=\mathcal{G}_\theta(\rho_s,\mu),\quad\mathcal{G}_\theta\approx e^{\Delta s\mathcal{L}},
\end{equation}
with the kinetic generator decomposed as
\begin{equation}
    \mathcal{L} = \mathcal{L}_{\mathrm{ext}} + \mathcal{L}_{\mathrm{wake}}.
\end{equation}
Using Lie--Trotter splitting,
\begin{equation}
    e^{\Delta s \mathcal{L}}\approx e^{\Delta s \mathcal{L}_ {\mathrm{wake}}} e^{\Delta s \mathcal{L}_{\mathrm{ext}}},
\end{equation}
which reflects the approximate factorization of collective beam transport into external lattice propagation and non-local wakefield interactions~\cite{hairer2006geometric}. This structure is particularly well suited for Fourier--Neural--Operator (FNO) based operator learning because collective effects enter through convolutional operators, as is the case for wakefields. In this case, Fourier representation becomes multiplicative, directly matching the spectral kernel formulation of FNO layers,
\begin{equation}
    \mathcal{K}_{\theta}(v)=\mathcal{F}^{-1}\!\left(R_{\theta}(k,\mu)\,\mathcal{F}(v)(k)\right).
\end{equation}
Consequently, the architecture naturally captures the non-local structure of impedance driven collective dynamics while remaining computationally efficient for long-range interactions and even multi-scale beam evolution~\cite{li2021fourier}. In this formulation machine and impedance parameters $ \mu=(I_{\mathrm{bunch}},a_{\mathrm{pipe}},Z_{\mathrm{scale}},\ldots)$ appear explicitly inside the spectral kernel $R_{\theta}(k,\mu)$, allowing the neural operator to learn entire parametric families of collective wake operators rather than a single fixed transport map~\cite{lu2021learning}. As a result, the model can generalize across varying accelerator configurations and beam conditions while preserving the underlying operator structure governing collective transport. Fig.~\ref{fig:fno1d} illustrates a representative example of longitudinal beam transport through a section of the HEB lattice together with the corresponding evolution of the beam distribution learned by the operator network.

\section{CONCLUSION}

An updated impedance and wake model has been developed for the FCC-ee HEB. The model includes the resistive wall contribution of the circular multilayer beam pipe and the geometric contributions from bellows, BPMs, and RF cavities. The resistive wall impedance is evaluated analytically, while the geometric contributions are obtained from CST simulations.

The present impedance and wake budgets are dominated by the resistive wall contribution. This indicates that the geometric contributions from the present component designs remain subdominant in the total budget. The RF cavity contribution nevertheless requires further investigation, since its relative impact appears larger than in the corresponding main ring studies.

The updated impedance and wake model will be used as input for future beam dynamics studies, including single bunch tracking, transverse mode coupling instability scans, and coupled bunch instability analyses.

\end{document}